\documentclass[fleqn,twoside]{article}
\usepackage{espcrc2}

\usepackage{graphicx}
\usepackage[figuresright]{rotating}

\newcommand{\AmS}{{\protect\the\textfont2
  A\kern-.1667em\lower.5ex\hbox{M}\kern-.125emS}}

\title{Light Bottom Squark Phenomenology}

\author{Edmond L. Berger
\address
{High Energy Physics Division, 
        Argonne National Laboratory,  
        Argonne, IL 60439, USA}%
        \thanks{I thank B.~Harris, C.-W.~Chiang, L. Clavelli, 
	J.~Jiang, D.~E.~Kaplan, J.~Lee, Z.~Sullivan, T.~Tait, and 
	C.~Wagner for their collaboration. Work in the High Energy Physics 
        Division at Argonne is supported by the U.S. Department of Energy, 
        Division of High Energy Physics, Contract W-31-109-ENG-38. }}
\begin{document}
\begin{abstract}
Agreement of theoretical calculations with the observed production rate of bottom 
quarks at hadron colliders is improved by the introduction of a contribution from 
pair-production of light gluinos, of mass 12 to 16 GeV, having two-body decays into 
bottom quarks and light bottom squarks with mass $\simeq 2$ to 5.5 GeV. 
Predictions are made for hadronic and radiative decays of the Upsilon states.  In the 
limit of large $\tan\beta$, the dominant decay mode of the light scalar Higgs boson 
is into a pair of light bottom squarks that materialize as jets of hadrons.      
\vspace{1pc}
\end{abstract}
%
% typeset front matter (including abstract)
\maketitle
\section{INTRODUCTION}
The cross section for bottom-quark $b$ production at hadron collider energies 
exceeds the central value of predictions of next-to-leading order (NLO) 
perturbative quantum chromodynamics (QCD) by about a factor of
two or three~\cite{expxsec}.  The NLO contributions are large, and a combination 
of further higher-order effects in production and/or fragmentation may eventually 
reduce the discrepancy~\cite{frag}.  In Ref.~\cite{Berger:2000mp}, my 
collaborators and I propose a contribution from physics beyond the standard model 
(SM).  In this paper, I summarize the proposal in Ref.~\cite{Berger:2000mp} of light 
gluinos $\tilde{g}$ and light bottom squarks $\tilde{b}$, as well as subsequent 
work~\cite{Berger:2001jb,Berger:2002kc,Berger:2002gu,Berger:higgs,Berger:raddecay}.
In this scenario the $\tilde b$ is the lightest supersymmetric (SUSY) particle, and 
the masses 
of all other SUSY particles are arbitrarily heavy, i.e., of order the electroweak 
scale or greater.  The lifetime of the $\tilde{b}$ is assumed to be less than 
the cosmological time scale so that these squarks make no contribution to the 
dark matter density. References to an extensive body of other recent theoretical 
papers can be found in~\cite{Berger:higgs}.  Various experimental constraints 
and phenomenological implications are examined in Ref.~\cite{Berger:2002kc}. 

There are important restrictions on couplings of the $\tilde{b}$ from precise 
measurements of $Z^0$ decays.  A light $\tilde b$ would be ruled out unless its 
coupling to the $Z^0$ is very small.  The squark couplings to the $Z^0$ depend 
on the mixing angle $\theta_b$. The lowest-order (tree-level) coupling to the $Z^0$ 
can be arranged to be small~\cite{light-sb} if $\sin^2 \theta_b \sim 1/6$.  The 
couplings of the heavier bottom squark ${\widetilde{b}_2}$ survive.  A careful 
phenomenological 
analysis is needed of expected $\widetilde{b}_2$ decay signatures, along with 
an understanding of detection efficiencies and expected event rates, before one 
knows the admissible range of its masses consistent with LEP data.  In the first paper 
of Ref.~\cite{Cao:2001rz} it is argued that one-loop contributions may render the 
light $\tilde{g}$ and light $\tilde{b}$ scenario inconsistent with data, unless the 
mass of the ${\widetilde{b}}_2$ is less than about 125 GeV.
The possibility that ${\widetilde{b}}_2$ lies in this mass range is not excluded.
In the second paper, the mass bound is relaxed to about 180 GeV,
and in the third, the constraint is further relaxed if $CP$-violating phases
are present. 
\section{HADRON COLLIDERS}
The light gluinos are produced in pairs via standard QCD subprocesses, dominantly 
$g + g \rightarrow \tilde g + \tilde g$ at Tevatron and Large Hadron Collider 
(LHC) energies.  The $\tilde g$ has
a strong color coupling to $b$'s and $\tilde b$'s and, as long as its mass
satisfies $m_{\tilde g} > m_b + m_{\tilde b}$, the $\tilde g$ decays
promptly to $b + \tilde b$.  The magnitude of the $b$ cross section, the
shape of the $b$'s transverse momentum $p_{Tb}$ distribution, and the CDF
measurement~\cite{cdfmix} of $B^0 - \bar B^0$ mixing are three features of
the data that help to establish the preferred masses of the $\tilde g$ and
$\tilde b$.  
Values of $m_{\tilde g} \simeq$ 12 to 16 GeV are chosen because the resulting 
$\tilde g$ decays produce $p_{Tb}$ spectra that are enhanced primarily in the
neighborhood of $p_{Tb}^{\rm min} \simeq m_{\tilde g}$ where the data show
the most prominent enhancement above the QCD expectation.  Larger values of
$m_{\tilde g}$ yield too little cross section to be of interest, and
smaller values produce more cross section than seems tolerated by the ratio
of like-sign to opposite-sign leptons from $b$ decay.

After the contributions of the NLO QCD and SUSY components are added, 
the magnitude of the bottom-quark cross section and the shape of the integrated 
$p^{\rm min}_{Tb}$ distribution are described well. 
The SUSY process produces $b$'s in a four-body final state.  Nevertheless, 
the angular correlations between $b$'s in the SUSY case are nearly 
indistinguishable from those of QCD once experimental cuts are applied.    
The energy dependence of the $b$ cross section is a potentially
important constraint on models in which new physics is invoked.  Since 
the assumed $\tilde{g}$ mass is larger than the mass
of the $b$, the $\tilde{g}$ pair process will turn on more slowly with
energy than pure QCD production of $b \bar{b}$ pairs.  The new physics
contribution will depress the ratio of cross sections at 630 GeV and
1.8 TeV from the pure QCD expectation.  An explicit calculation with 
CTEQ4M parton densities and the $b$ rapidity selection $|y| < 1$, yields 
a pure NLO QCD prediction for the ratio of 0.17 +/- 0.02 for 
$p_{Tb}^{\rm min} =$ 10.5 GeV, and 0.16 +/- 0.02 
after inclusion of the $\tilde{g}$ pair contribution.  Either of these numbers 
is consistent with data~\cite{Acosta:2002qk}.   
\subsection{Like/Unlike-sign B's and Leptons}
If, as in many scenarios, the $\tilde g$ is a Majorana particle, its decay 
yields both quarks and antiquarks.  Pair production of Majorana gluinos and 
subsequent decay to $b$'s will generate $b b$ and $\bar b \bar b$ pairs, as well 
as the $b \bar b$ final states that appear in QCD production.  Therefore, a 
``gold-plated'' prediction is production of $B^+ B^+$ and $B^-B^-$ pairs.  
For the cuts chosen in current hadron collider experiments, an equal number of 
like-sign and opposite-sign $b$'s is expected from the SUSY 
mechanism, leading to an increase of like-sign leptons in the final state after 
semi-leptonic decays of the $b$ and $\bar b$ quarks.  This increase could be 
confused with an enhanced rate of $B^0-\bar B^0$ mixing.  

Time-integrated mixing analyses of lepton pairs 
observed at hadron colliders are interpreted in terms of the 
quantity $\bar{\chi}$. 
The CDF measurement~\cite{cdfmix} of 
$\bar{\chi}_{\rm {eff}} = 0.131 \pm 0.02 \pm 0.016$   
is marginally larger than the world average value $\bar{\chi} = 
0.118 \pm 0.005$~\cite{pdg}, assumed to be the contribution from the pure QCD 
component only.  After the contribution from new physics is included, the 
predictions are $\bar{\chi}_{\rm {eff}} = 0.17 \pm 0.02 $ for 
$m_{\tilde g} =$ 14 GeV, and 
$\bar{\chi}_{\rm {eff}} = 0.16 \pm 0.02 $ with $m_{\tilde g} =$ 16 GeV.  The 
calculated $\bar{\chi}_{\rm {eff}}$ is consistent with the data within 
uncertainties if $m_{\tilde g} > 12$ GeV.  The published result is based on 
an analysis of only 20\% of the run-I sample and only the $\mu \mu$ final 
state.  It would be valuable to extend the analysis to the full sample in both 
the $e \mu$ and $\mu \mu$ modes.  

With $\sigma_{\tilde{g}\tilde{g}} / \sigma_{\rm{qcd}} \sim 1/3$, 
the mixing data and the magnitude and $p_T$ dependence of the $b$ production 
cross section can be satisfied.  
\section{$\Upsilon$ DECAY}
If $m_{\tilde{b}}$ is less than half the mass of one of the Upsilon states,  
then $\Upsilon$ decay to a pair of bottom squarks might proceed with sufficient rate 
for observation or exclusion of a light $\tilde{b}$.  The 
rate for $\Upsilon(nS) \rightarrow \tilde b {\tilde b}^*$ is computed in 
Ref.~\cite{Berger:2001jb} as a function of the masses of the $\tilde{b}$ and 
the $\tilde{g}$, and $\chi_{bJ}$ decays are treated in Ref.~\cite{Berger:2002gu}.      
The data sample is largest at the $\Upsilon(4S)$.  For a fixed 
$\tilde{g}$ mass of 14 GeV, the branching fraction into a pair of $\tilde{b}$'s
is about $10^{-4}$ for $m_{\tilde b} =$ 4.85 GeV.  A large sample 
may be available from the CLEO, BaBar, and BELLE experiments. 
Direct observation 
of $\Upsilon(nS)$ or $\chi_b$ decay into $\tilde{b}$'s requires an understanding of the 
ways that $\tilde{b}$'s may manifest themselves, discussed in 
Refs.~\cite{Berger:2001jb,Berger:2002kc,Berger:2002gu}.  
Possible baryon-number-violating R-parity-violating decays of the $\tilde{b}$ lead to 
$u+s$; $c+d$; and $c+s$ final states.  

It is possible that the $\tilde{b}$ 
is relatively stable and, hence, bound states of a bottom 
squark and bottom antisquark (sbottomonium) could exist. These bound states 
could be produced in radiative decays of bottomonium states, such as 
$\Upsilon \rightarrow \tilde{S} \gamma$, where $\tilde{S}$ is the $S$-wave 
bound state of a $\tilde{b} \tilde{b}^*$ pair.  In Ref.~\cite{Berger:raddecay}, 
a calculation is presented of the radiative decay of the $\Upsilon(nS)$
states into a bound state of $\tilde{b}$'s.  Predictions are provided of the
branching fraction as a function of the masses $m_{\tilde{b}}$ and
$m_{\tilde{g}}$.  Branching fractions as large as several times $10^{-4}$ are
obtained for SUSY particle masses in the range suggested by
the analysis of the $b$ cross section.  Provided that a bound 
state can be formed, the resonance search by the CUSB 
Collaboration~\cite{CUSB-gamma} raises the allowed lower bounds on 
$m_{\tilde{b}}$ and $m_{\tilde{g}}$.  Discovery of the $\tilde{S}$ bound states 
may be 
possible with the high-statistics 2002 CLEO-c data set, or a larger 
range of $\tilde{b}$ and $\tilde{g}$ masses may be 
disfavored~\cite{Berger:raddecay}.
\section{HIGGS BOSON DECAY}
Current strategies for discovery and measurement of the properties the neutral 
scalar Higgs particle $h$ with $m_h < 135$ GeV rely heavily on the 
presumption that the principal branching fractions are close to those predicted 
in the SM or in the usual minimal supersymmetric standard model 
(MSSM).  For masses in this range, the decay width of 
the SM Higgs boson is dominated by its decay into bottom quarks, $b \overline{b}$. 
In Ref.~\cite{Berger:higgs}, my collaborators and I show explicitly that these 
assumptions are not warranted if there are non-standard light particles such as 
$\tilde{b}$'s in the spectrum.  We analyze the possibility that the $h$ 
decays into new particles that manifest themselves as hadronic 
jets without necessarily significant bottom or charm flavor content.  As an
example of this possibility, we present the case of a light $\tilde{b}$, with 
mass smaller than about 10 GeV.  

We work in the decoupling limit in which the mass of the pseudo-scalar 
Higgs boson ($m_A$) is large compared to $m_Z$, and we assume that 
the ratio of Higgs vacuum expectation values $\tan \beta$ is large.  
No assumption is made about the gluino mass; a light $\tilde{g}$ is not 
required.  Under these conditions, the dominant decay of $h$ is into a pair 
of light $\tilde{b}$'s.  The total decay width of the $h$ becomes several 
orders of magnitude larger than the width for decay into $b$'s. 
Branching fractions into SM decay channels are reduced from their 
SM values by a factor proportional to $\tan^{-2} \beta$.  
For values of the branching ratio $BR(h \to \tilde{b} \tilde{b}^*)$
larger than two to five times that into bottom quarks, the large QCD jet 
backgrounds will make observation of the $h$ very difficult in Tevatron and 
LHC experiments.  
Because they rely principally on the production process $e^+ e^- \rightarrow 
h Z^0$, experiments at proposed $e^+ e^-$ linear 
colliders remain fully viable for direct observation of the $h$ and 
measurement of its mass and some of its branching fractions~\cite{Berger:higgs}.  

\end{document}